\documentclass[prd,preprint,showpacs,preprintnumbers,nofootinbib,eqsecnum,superscriptaddress]{revtex4}

 \usepackage[dvips,final]{graphicx}
  \usepackage{amssymb}
   \usepackage{amsmath}
    \usepackage{amsfonts}
     \usepackage{epsfig}
      \usepackage{bm}

\usepackage{mathpazo}

\usepackage[section]{placeins}

\usepackage{multirow}
\usepackage{ctable}
\usepackage{booktabs}
\usepackage{array}
\usepackage{tabularx}
\usepackage{xcolor}
\usepackage{pstricks}

\begin{document}

\title{Search for walking technipion in four-jet sample}

\author{Rafa{\l} Maciu{\l}a}
\email{rafal.maciula@ifj.edu.pl} \affiliation{Institute of Nuclear
Physics, Polish Academy of Sciences, ul. Radzikowskiego 152, 
PL-31-342 Krak{\'o}w, Poland}

\author{Antoni Szczurek\footnote{also at University of Rzesz\'ow, PL-35-959 Rzesz\'ow, Poland}}
\email{antoni.szczurek@ifj.edu.pl} \affiliation{Institute of Nuclear
Physics, Polish Academy of Sciences, ul. Radzikowskiego 152, 
PL-31-342 Krak{\'o}w, Poland}

\date{\today}

\begin{abstract}
We discuss how to search for a possibe signal of the recently
observed 750 GeV enhancement in the diphoton channel in the four-jet production. In the present studies we assume that the produced
state is pseudoscalar. This fact, when combined with specificity of 
the corresponding amplitude, allows to improve the signal-to-background (S/B)
ratio. We discuss in detail how to impose
cuts on jets in rapidity and transverse momenta in order
to find optimal S/B ratio and not to loose
too much statistics.
Our study suggest a measurement of two soft (low cut on $p_t$) 
large-rapidity jets and two hard (high cut on $p_t$) mid-rapidity jets. 
Azimuthal correlation between the soft external
jets may be useful to further improve the situation.
Several differential distributions in rapidities and transverse momenta
of jets as well as dijet invariant mass are shown. 
The integrated cross sections corresponding to different cuts 
are collected in a table and number of events are presented.
\end{abstract}


\maketitle

\section{Introduction}

Last year the ATLAS and CMS collaborations observed an enhancement of the cross section 
at 750 GeV in the diphoton channel at 
$\sqrt{s}$ = 13 TeV \cite{ATLAS_13TeV,CMS_13TeV}.
Very recently also data searching for the potential signal 
at $\sqrt{s}$ = 8 TeV were released \cite{ATLAS_8TeV,CMS_8TeV}.
Different models were discussed recently in the context of 
the newly observed enhancement. A short review on the topic can 
be found in Ref.~\cite{Strumia_review_2016}. 
If the signal is true we know that the new object decays (couples) to 
photons (the observation channel).
The production mechanism is, however, not clear at present and we have
no any clear hint for this.
In some of the models considered recently the object is produced
dominantly by the gluon-gluon fusion \cite{Matsuzaki}, 
in some models dominantly by the photon-photon fusion 
\cite{Royon,LLPS2016}.
If the gluon-gluon fusion dominates the resonance should have large
width. In contrast, when photon-photon fusion dominates the resonance
should be narrow, much narrower than experimental resolution
in the diphoton invariant mass.
The recent analysis \cite{Strumia_review_2016} suggests that 
the gluon-gluon fusion may be preferred from the analysis of the ratio of 
the $\sqrt{s} =$ 8 TeV and $\sqrt{s} =$ 13 TeV cross sections.

In Ref.~\cite{LLPS2016} one of us discussed production of the signal
in a vector-like SU(2) technipion model as well as in a walking technipion
model \cite{Matsuzaki}. In the first case the signal is produced
mainly via photon-photon fusion while in the second one dominantly via
gluon-gluon fusion.
If models with gluon-gluon fusion (like-walking technipion model) 
are right then the object could be also potentially observed 
(or at least could be searched for) in the digluon (dijet) final state.
However, as shown in Ref.~\cite{LLPS2016}, the dijet background
makes practically impossible such an observation in the dijet invariant mass. 
Possible specificity of the couplings involved in the amplitude 
may cause that the signal is in large fraction of cases produced 
in association with one or two jets \cite{LLPS2016}.

\begin{figure}[!h]
\includegraphics[width=4cm]{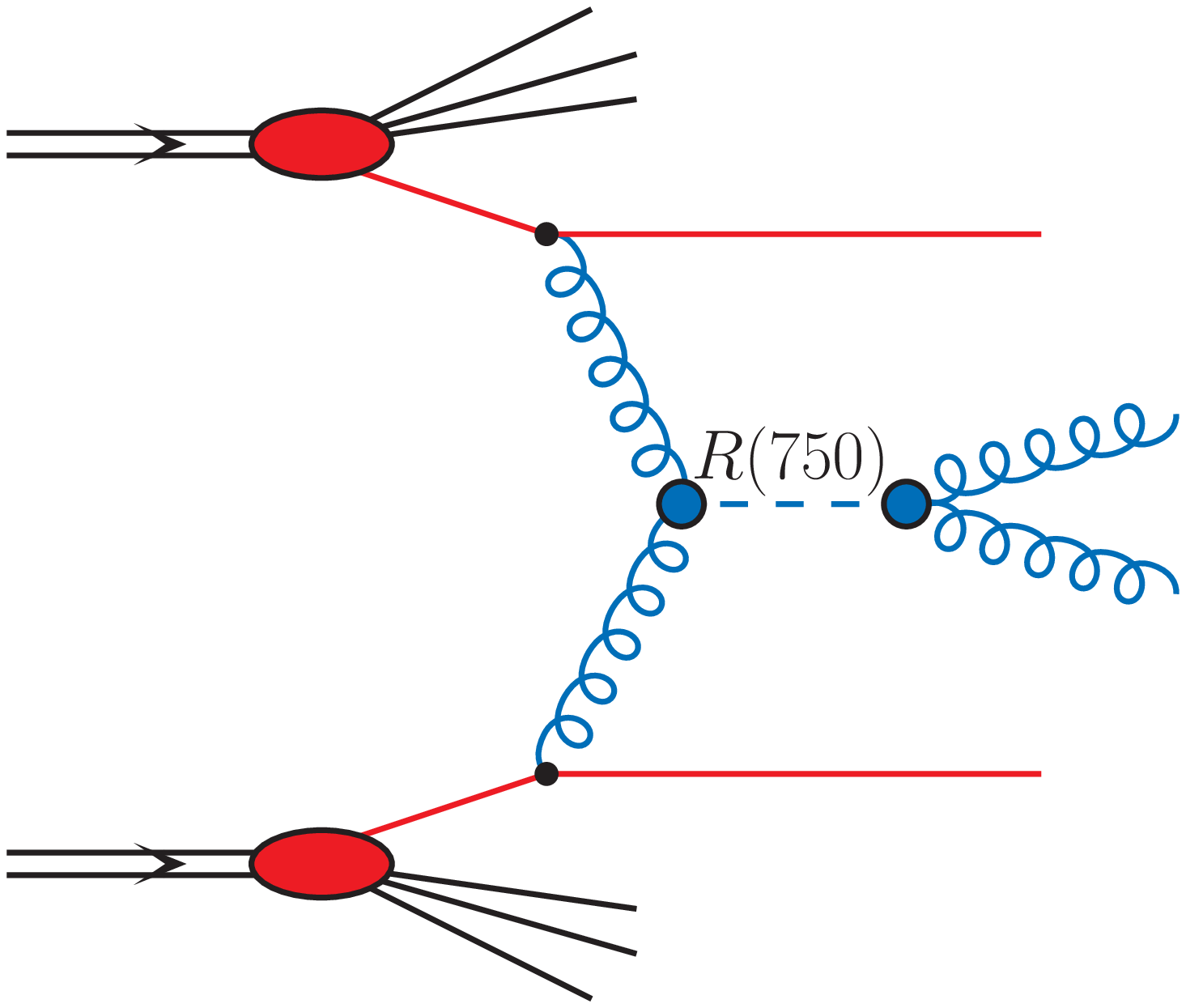}
\includegraphics[width=4cm]{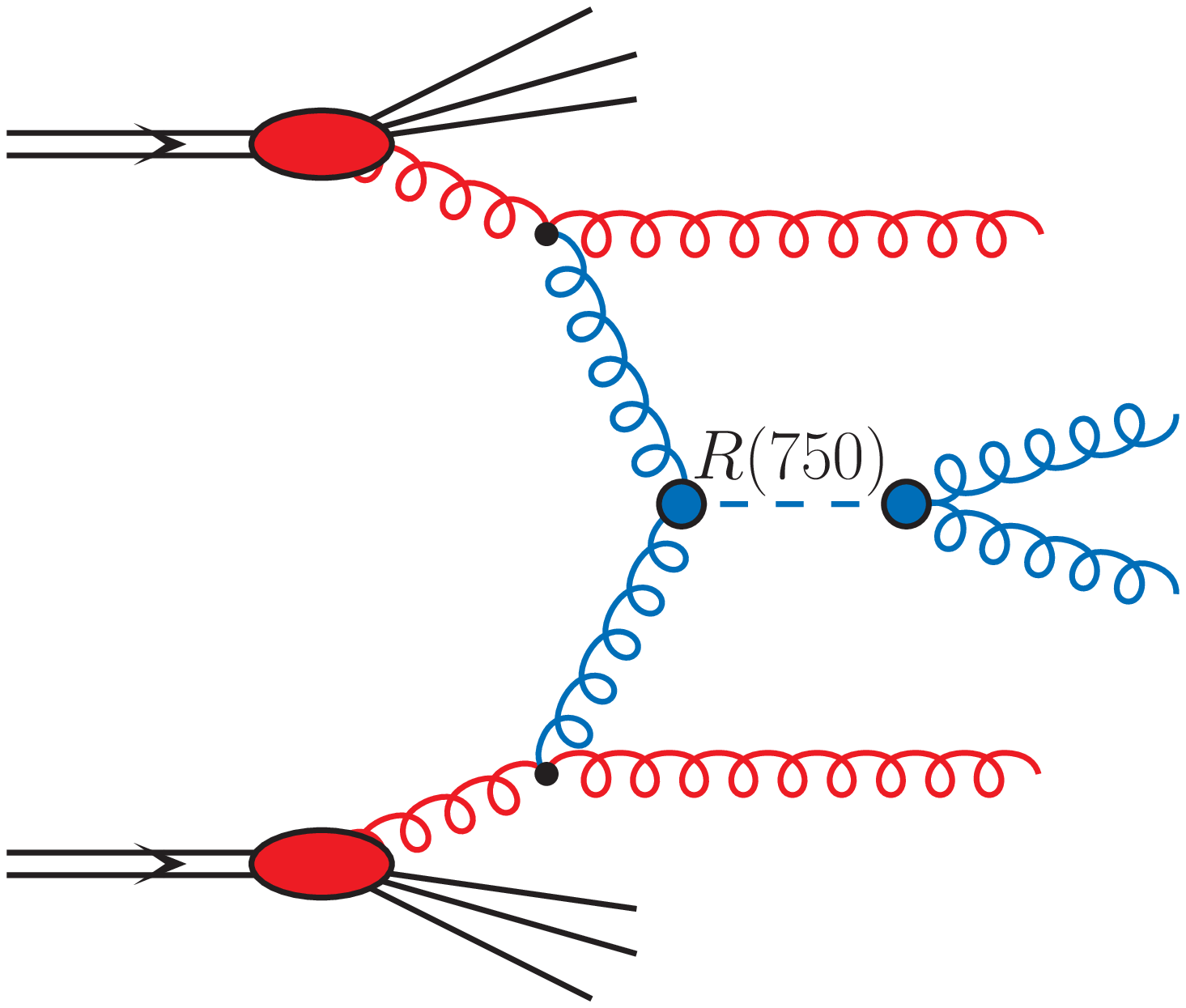}
\includegraphics[width=4cm]{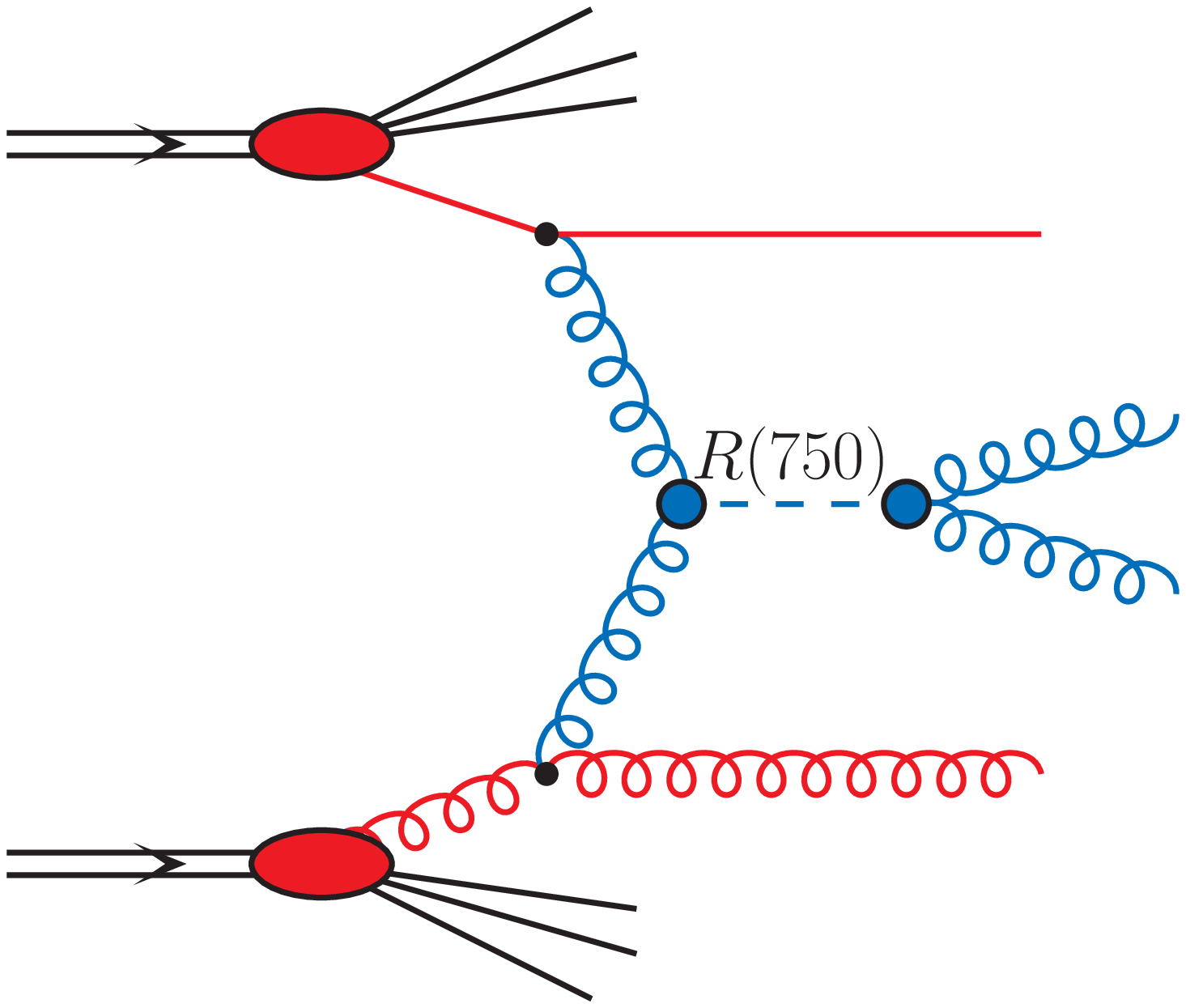}
\includegraphics[width=4cm]{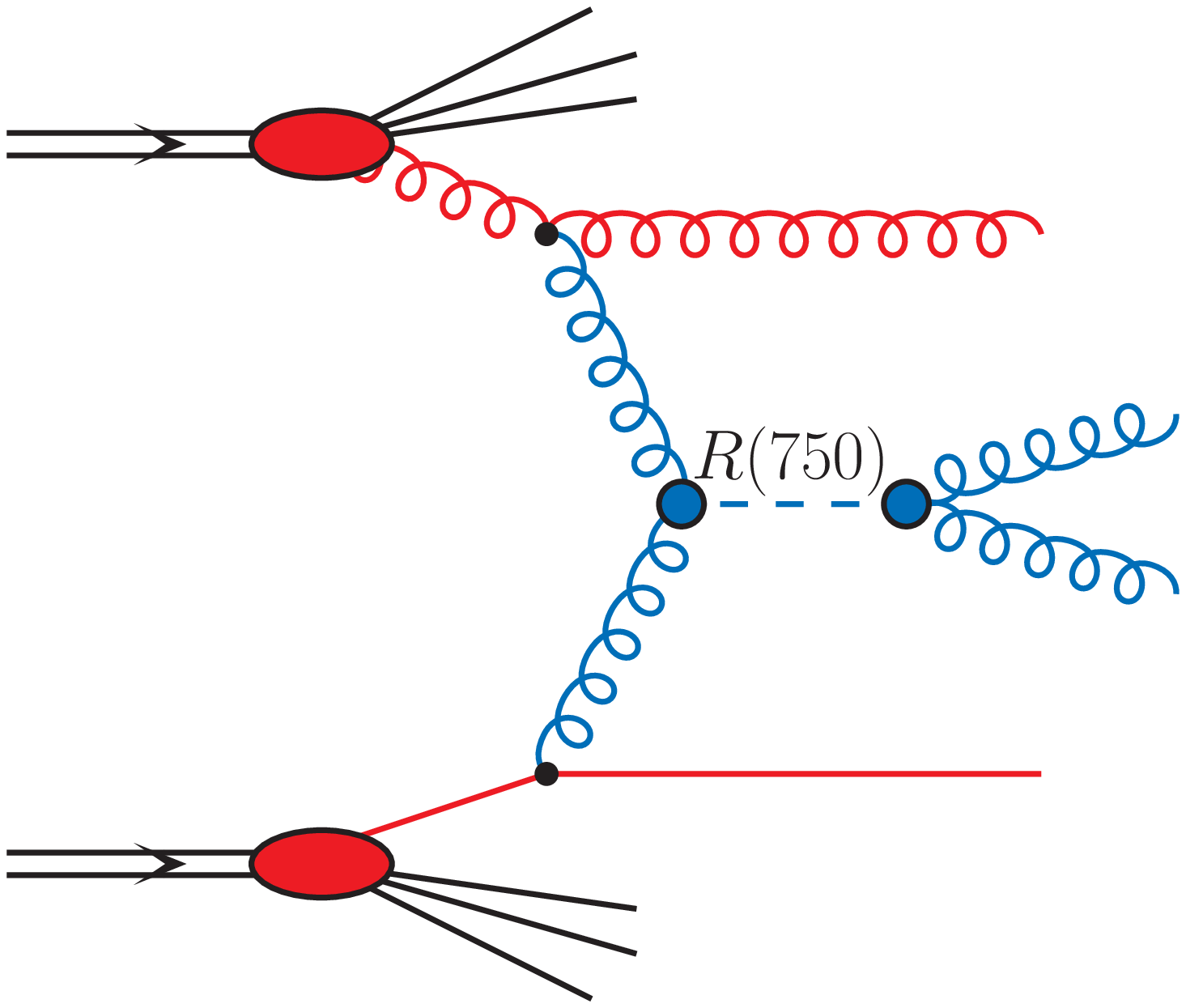}
   \caption{
\small The mechanisms of four-jet production with one intermediate
resonant state $R(750)$.
 }
 \label{fig:diagrams}
\end{figure}

Here we discuss how to search for the state in 
the four-jet sample
(the relevant diagrams for the signal are shown in Fig.~\ref{fig:diagrams}).
In the present case we consider production of the pseudoscalar object 
($J^{\pi} = 0^-$) discussed in detail in Ref.~\cite{LLPS2016}.

\section{A sketch of the theoretical formalism}

\subsection{Signal in four-jet final state}

When calculating signal we consider a simple $2 \to 3$
partonic subprocesses shown in Fig.~\ref{fig:diagrams}. Here for example we discuss only the first diagram.
The cross section for the partonic $q q' \to q {\tilde \pi}^0 q'$ 
process reads:
\begin{equation}
\sigma_{q q' \to q {\tilde \pi}^0 q'} = \frac{1}{2 {\hat s}}
\overline{ | {\cal M}_{q q' \to q {\tilde \pi}^0 q'}|^2 } {\cal J}
d \xi_1 d \xi_2 d y_{\tilde{\pi}^{0}} d \phi_{12} \,,
\end{equation}
where $\phi_{12}$ is the relative azimuthal angle between $q$ and $q'$,
$\xi_1 = \log_{10}(p_{1t})$ and $\xi_2 = \log_{10}(p_{2t})$,
where $p_{1t}$ and $p_{2t}$ are transverse momenta of outgoing
$q$ and $q'$, respectively.

The matrix element for the $2 \to 3$ subprocess is calculated as:
\begin{eqnarray}
&&{\cal M}_{q q' \to q {\tilde \pi}^0 q'}
(\lambda_1, \lambda_2, \lambda_3, \lambda_4) = e^{2}\,
\bar{u}(p_{3},\lambda_{3}) \gamma^{\mu} u(p_{1},\lambda_{1}) \,
\frac{-i g_{\mu \nu}}{\hat{t}_{1}}\nonumber\\
&&\qquad \qquad \qquad \times 
\varepsilon^{\nu \nu' \alpha \beta} q_{1\alpha} q_{2\beta}\,
F_{g g \to R}
\frac{-i g_{\nu' \mu'}}{\hat{t}_{2}}\,
\bar{u}(p_{4},\lambda_{4}) \gamma^{\mu'} u(p_{2},\lambda_{2})\,.
\label{ME_2to3}
\end{eqnarray}
The $F_{g g \to R}$ effective coupling, except of normalization,
is the same as the $F_{\gamma \gamma \to R}$ one discussed
in detail in Ref.~\cite{LLPS2016}.
As in Ref.~\cite{LLPS2016}, the normalization is adjusted to experimental data on diphoton
production using branching fractions as obtained in the
walking technicolor model \cite{Matsuzaki}.

The matrix elements for the other processes in Fig.~\ref{fig:diagrams} 
can be obtained easily in the high-energy approximation by replacing
appropriate coupling constants and color factors 
(see e.g. \cite{SLM2014}). As in Ref.~\cite{LLPS2016} for comparison, we also calculated the matrix 
element in the high-energy approximation:
\begin{eqnarray}
\bar{u}(p', \lambda') \,
\gamma^{\mu} \,
u(p, \lambda) 
\rightarrow 
(p' + p)^{\mu}\, 
\delta_{\lambda' \lambda}\,.
\label{ME_2to3_he}
\end{eqnarray}
We have also obtained a formula for matrix element squared with
algebraic computer calculation and checked
that it gives the same result as the calculation with explicit
use of spinors.

In order to impose cuts on transverse momenta and rapidity of jets
we calculate first the distribution 
\begin{equation}
\frac{d \sigma}{d y_R d p_{t,R}}(y_R, p_{t,R}; \text{with explicit cuts on spectator jets}).
\end{equation}
Then the decay of $R \to g g$ is done in a separate simple Monte Carlo
code (four-momenta of gluons are obtained) 
where cuts on gluons from the decay of the resonance are imposed
in addition.

\subsection{Four-jet background}

Four-jet production via single-parton scattering (SPS) mechanism in proton-proton scattering at the LHC was discussed
some time ago in the collinear factorization, both at leading-order (LO) and, for the first time, at next-to-leading order (NLO) in Ref.~\cite{Bern:2011ep}.
Then, very recently also first four-jet studies based on the tree-level $k_{T}$-factorization approach with two off-shell partons have been performed, including double-parton scattering (DPS) effects \cite{Kutak:2016mik}. In the present paper we follow the LO collinear approach in order to make the background calculations consistent with the predictions for the signal, so the S/B ratio can be estimated more or less in a model independent way.

According to the LO collinear approach the elementary cross section for the SPS mechanism of four-jet production
has the following generic form:
\begin{equation}
d\hat{\sigma}_{ij \to klmn} = \frac{1}{2\hat{s}} \; 
\overline{|{\cal M}_{i j \rightarrow k l m n}|^2} \; d^{4} PS ,
\label{elementary_cs}
\end{equation}
where the invariant phase space reads:
\begin{equation}
d^{4} PS = \frac{d^3 p_1}{2 E_1 (2 \pi)^3} \frac{d^3 p_2}{2 E_2 (2 \pi)^3}
           \frac{d^3 p_3}{2 E_3 (2 \pi)^3} \frac{d^3 p_4}{2 E_4 (2 \pi)^3}
           (2 \pi)^4 \delta^4 \left( p_1 + p_2 + p_3 + p_4 \right) \; . 
\label{four_body_phase_space}
\end{equation}
Above $p_1, p_2, p_3, p_4$ are four-momenta of outgoing partons (jets) and
${\cal M}_{i j \rightarrow k l m n}$ is the relevant $2 \to 4$ tree-level on-shell matrix element.

The hadronic cross section is then given by the integral
\begin{eqnarray}
d \sigma &=& \int d x_1 d x_2 \sum_{ijklmn}
           f_i(x_1,\mu_F^2) f_j(x_2,\mu_F^2) 
           \; d \sigma_{ij \to k l m n} 
      \; .
\label{hadronic cross section}
\end{eqnarray}
Above $f_i$ and $f_j$ are the standard collinear parton distribution functions (PDFs).

Instead of explicitly using the formulae above we shall use
its Monte Carlo version as implemented in the event generator ALPGEN
\cite{Mangano:2002ea} that is dedicated to multi-parton production studies.
In the numerical calculations here we are working in the $n_{F}=3$ flavour scheme and use a running $\alpha_{S}$ as implemented in the ALPGEN code
and the MSTW2008nlo68cl PDF sets \cite{Martin:2009iq}. We set both the renormalization and factorization scales
equal to the sum of the final state transverse momenta squared, $\mu_{R} = \mu_{F} = \sum_{i} p_{i,t}^{2}$.
The LO results for the background are corrected in the last step by the relevant $K = 0.5$ factor, which is taken from the
comparison of the LO and NLO collinear predictions with the multi-jet ATLAS data \cite{Bern:2011ep}. Recently, the similar comparison has been done, but for the tree-level $k_{T}$-factorization calculations and CMS data, and it seems to confirm the importance of the $K$-factor in the case of four-jet studies \cite{Kutak:2016ukc}.

\section{Numerical feasibility studies}

The decay of (pseudo)scalar R(750) \footnote{This statement
is universal, independent of the observed 750 GeV enhancement, however the
measured signal for the 750 GeV enhancement is rather unnaturally
large.} leads dominantly to not too soft (small transverse momentum) jets.
On the other hand to improve statistics the other two jets can (should) 
be as soft as possible. As far as the signal is considered we shall call
the jets from the decay of the R(750) resonance as ,,internal'' and 
the other two associated jets as ,,external''. 
These names are related to the position in diagrams and in rapidity space.
As minimal set of cuts we select two leading jets with $p_t >$ 100 GeV
and the two other jests with $p_t >$ 20 GeV. The leading jets are
assumed to be within the main (CMS or ATLAS) detector and the subleading
jets to be in the (CMS or ATLAS) calorimeters.

\begin{figure}[!h]
\begin{minipage}{0.47\textwidth}
 \centerline{\includegraphics[width=1.0\textwidth]{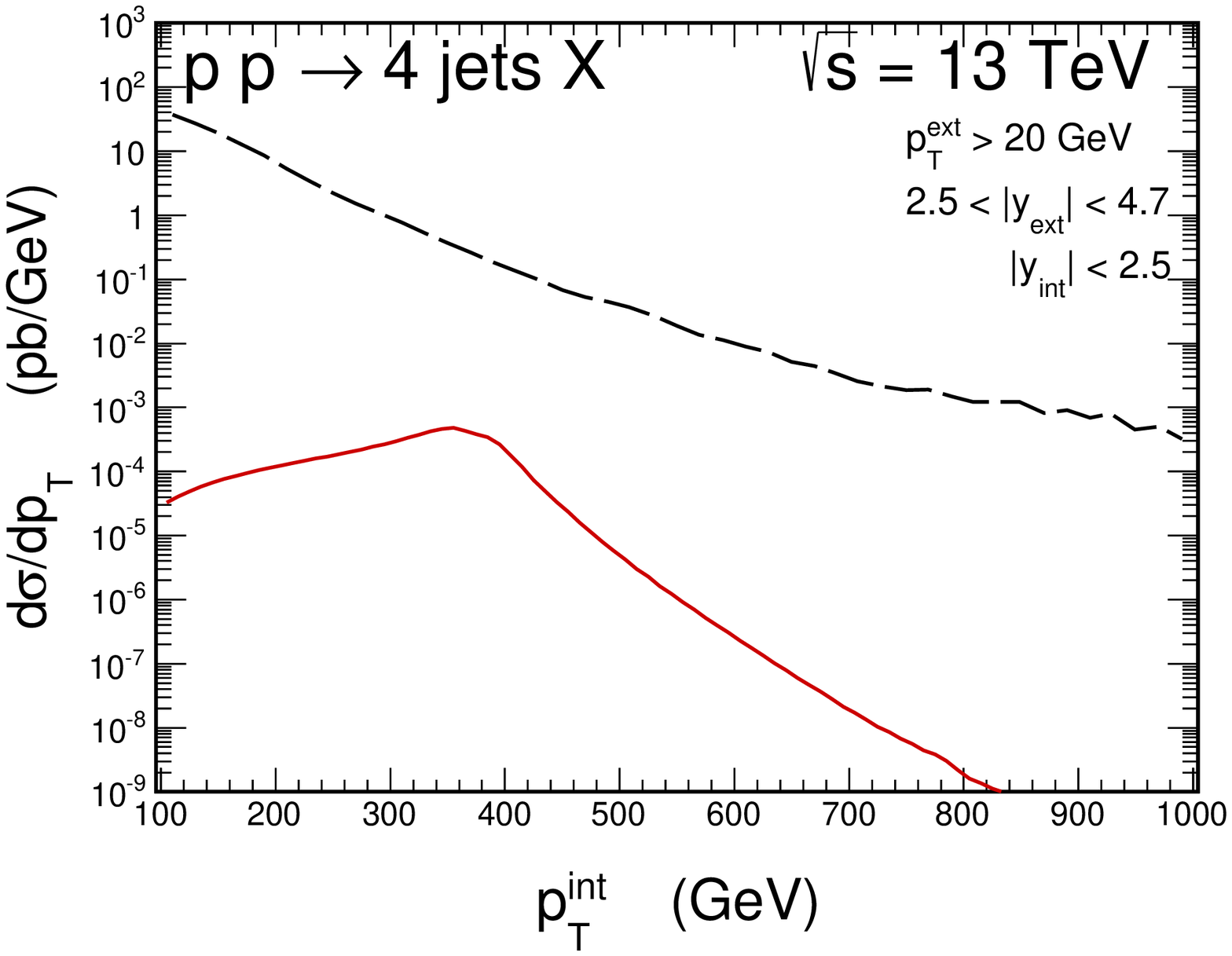}}
\end{minipage}
\hspace{0.5cm}
\begin{minipage}{0.47\textwidth}
 \centerline{\includegraphics[width=1.0\textwidth]{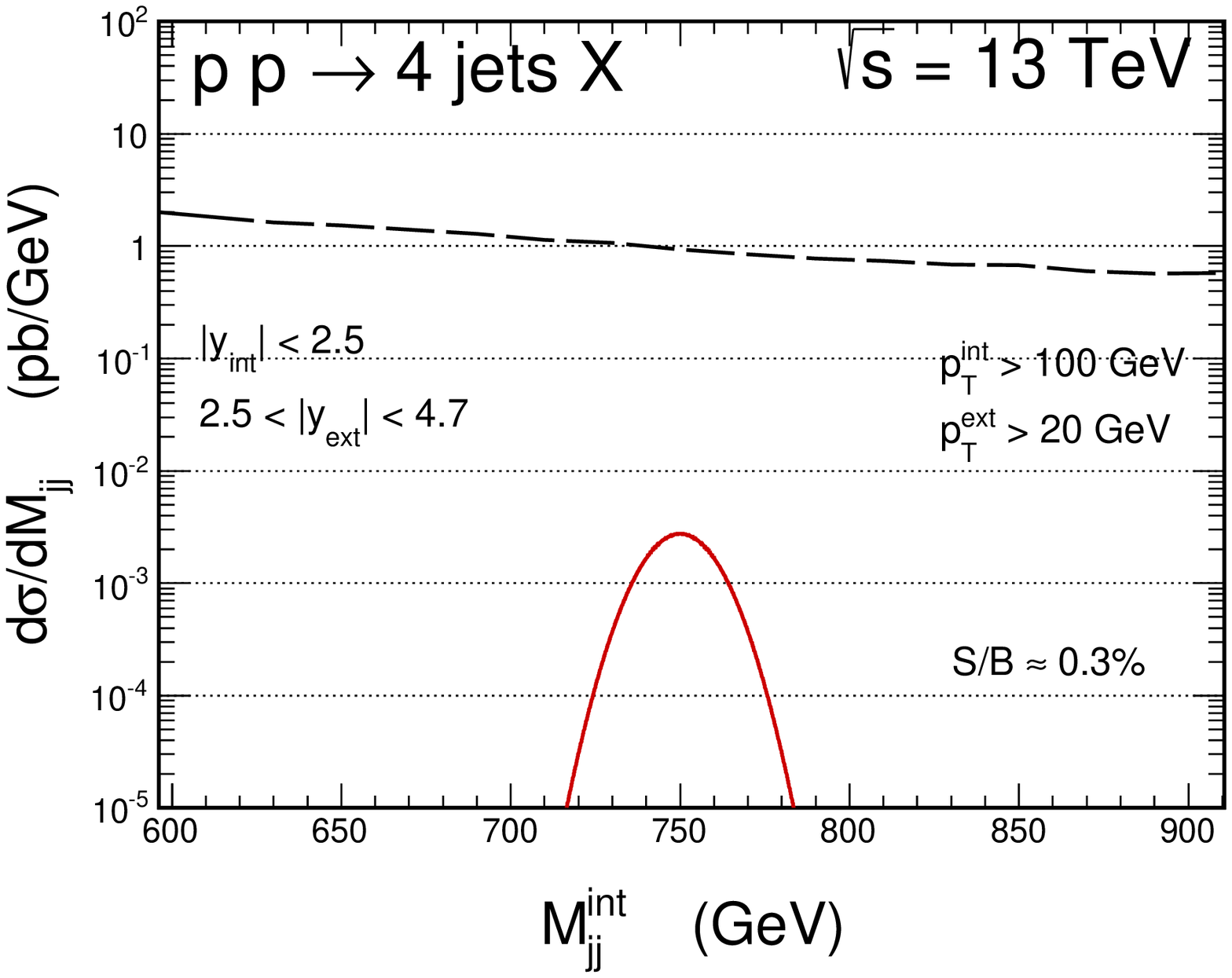}}
\end{minipage}
   \caption{
{\small Distribution in the transverse momentum of the hard-internal jets (left
panel) and in the invariant mass of the two hard-internal jets (right panel).
 }
 }
 \label{fig:pT_Minv_full}
\end{figure}

In Fig.~\ref{fig:pT_Minv_full} we show the corresponding cross sections
in jet transverse momentum (left panel) and dijet invariant mass (right panel) of the hard jets (internal jets for the
signal). Here we show both signal (solid, red on-line) and background (dashed)
distributions.
One can observe that with this minimal cut the signal is much below 
the four-jet background. The distribution in transverse momentum
of the hard-internal jets suggests that they should be selected with transverse momenta arround $M_R/$ 2,
where $M_{R}$ is the mass of the resonance (in the considered here case $M_{R} = 750$ GeV). 
Calculating invariant mass distribution we have assumed that the actual
total decay width is smaller than experimental dijet invariant mass 
resolution.
For the experimental resolution we assumed here Gaussian distribution
with width $\sigma =$ 10 GeV 
(see discussion in Ref.~\cite{LLPS2016}).

\begin{figure}[!h]
\begin{minipage}{0.47\textwidth}
 \centerline{\includegraphics[width=1.0\textwidth]{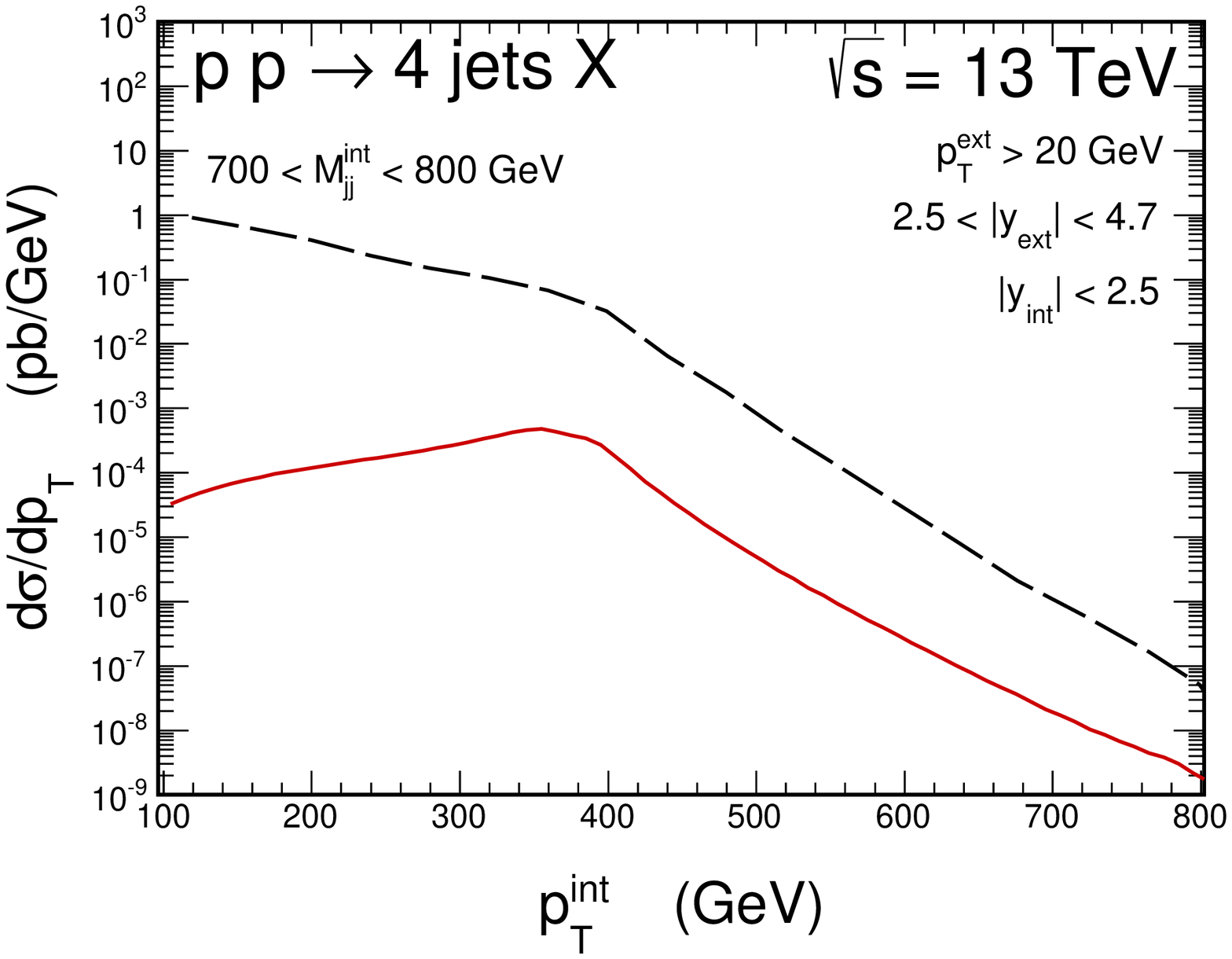}}
\end{minipage}
\hspace{0.5cm}
\begin{minipage}{0.47\textwidth}
 \centerline{\includegraphics[width=1.0\textwidth]{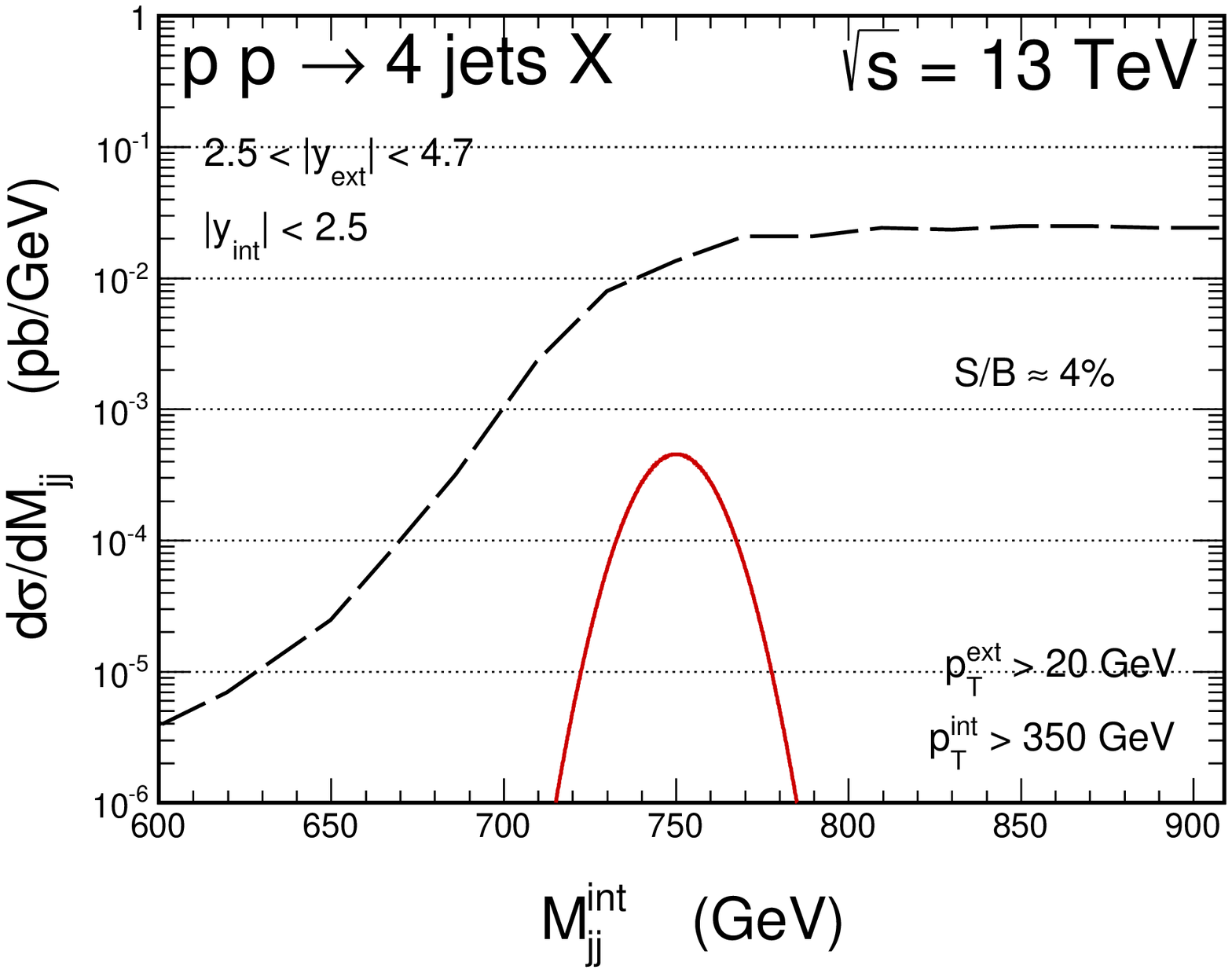}}
\end{minipage}
   \caption{
{\small Distribution in the transverse momentum of the hard jets with
invariant mass window (specified in the figure legend) around the signal
(left panel) and in the invariant mass of the two hard jets
with the increased lower cut on transverse momenta of hard internal jets (right panel).
 }
 }
 \label{fig:pT_Minv_window}
\end{figure}

The distribution in transverse momentum of the hard-internal
jets for the background is a bit misleading because it cannot
be directly compared with the signal. Therefore in 
the left panel of Fig.~\ref{fig:pT_Minv_window} we show similar distribution 
but now in much narrower window of dijet invariant mass 
(700 GeV $< M_{jj} <$ 800 GeV). Now the signal-to-background (S/B) ratio
looks much better, especially for $p_{T}^{int} >$ 350 GeV. 
Therefore in the right panel of Fig.~\ref{fig:pT_Minv_window} we show
invariant mass distribution imposing extra cut on the internal jets $p_{T}^{int} >$ 350 GeV.
Now the signal is still more than order of magnitude below the
background. Can we improve the situation even more?
Below we shall try to include specific features of the considered 
model amplitude for the signal.

\begin{figure}[!h]
\begin{minipage}{0.47\textwidth}
 \centerline{\includegraphics[width=1.0\textwidth]{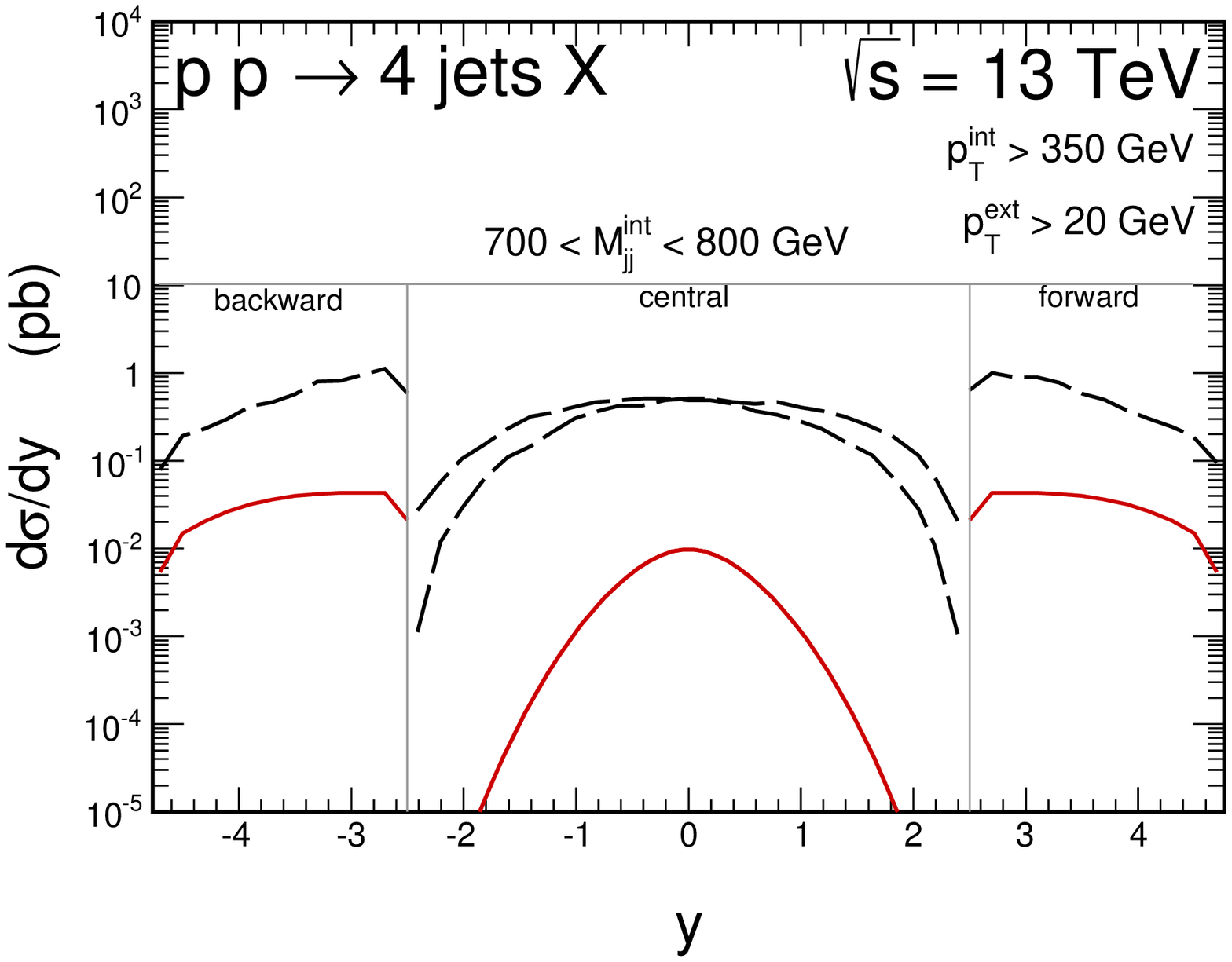}}
\end{minipage}
\hspace{0.5cm}
\begin{minipage}{0.47\textwidth}
 \centerline{\includegraphics[width=1.0\textwidth]{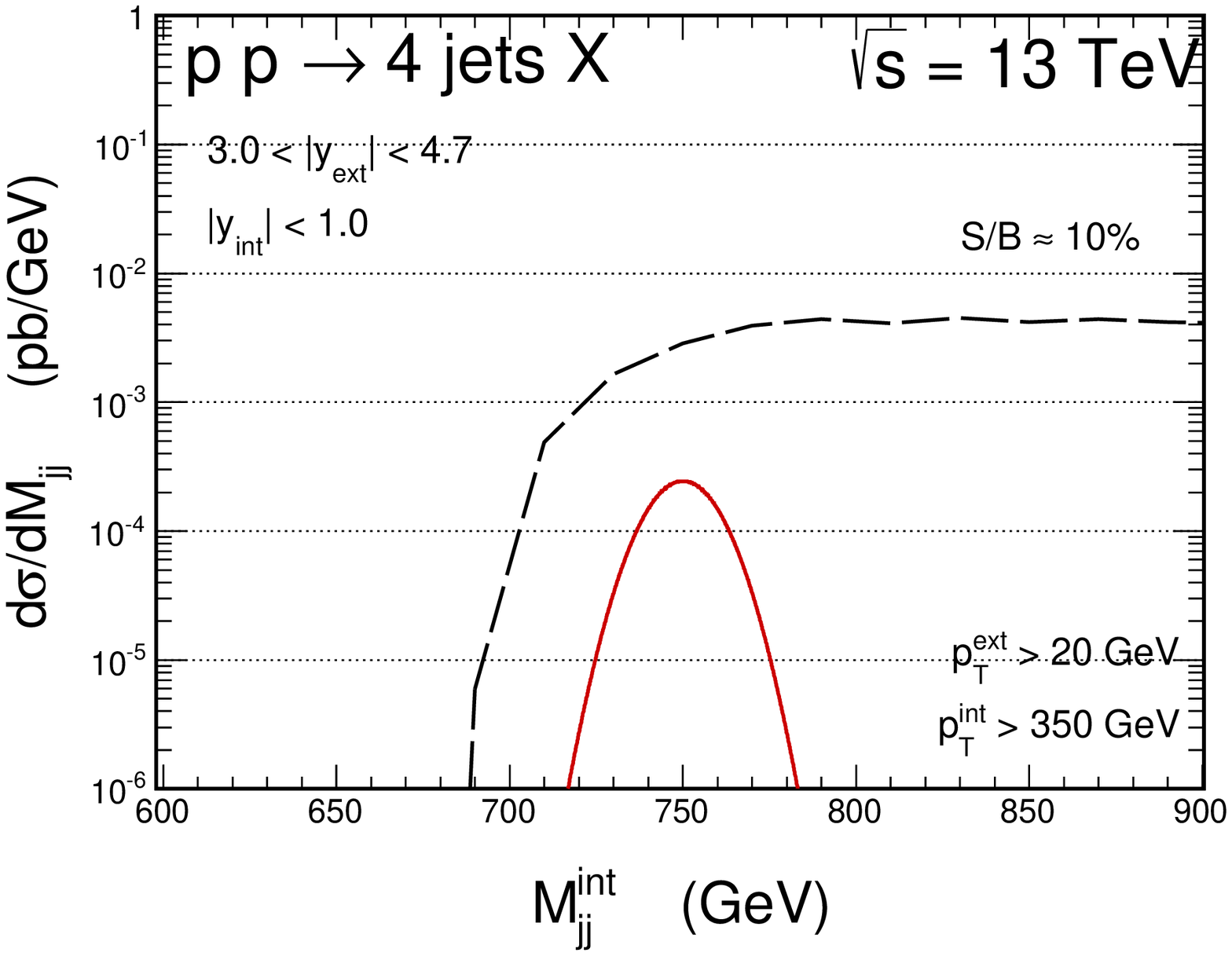}}
\end{minipage}
   \caption{
{\small Distribution in rapidities of internal and external jets
with the selected invariant mass window around the signal and for
increased lower cut on the transverse momenta of internal jets (left
panel) and the corresponding distribution in internal dijet invariant
mass with the optimized rapidity cuts specified in the figure legend (right panel).
} }
 \label{fig:y_Minv_window}
\end{figure}

So far we have taken only a rough choice of preferable cuts on jet
rapidities as dictated by pseudorapitity limitations of the tracker ($|y| < 2.5$) 
and calorimeters ($2.5 < |y| < 4.7$). 
Can the S/B ratio be further improved by a better choice 
of the rapidity cuts? The situation
is illustrated in the left panel of Fig.~\ref{fig:y_Minv_window}.
The plot shows that the jets from the decay of the R(750)
resonance are centered at midrapidities. In addition, one could
take larger rapidity cuts for the soft-external jets.
Our detailed study has shown that the rapidity cuts $|y| < 1$ for the hard-internal and $3 < |y| < 4.7$ for the soft-external jets are optimal as far as signal-to-background ratio and statistics
is considered. Now the S/B ratio of about~0.1 can be
achieved that is already a significant improvement.

\begin{figure}[!h]
\begin{minipage}{0.47\textwidth}
 \centerline{\includegraphics[width=1.0\textwidth]{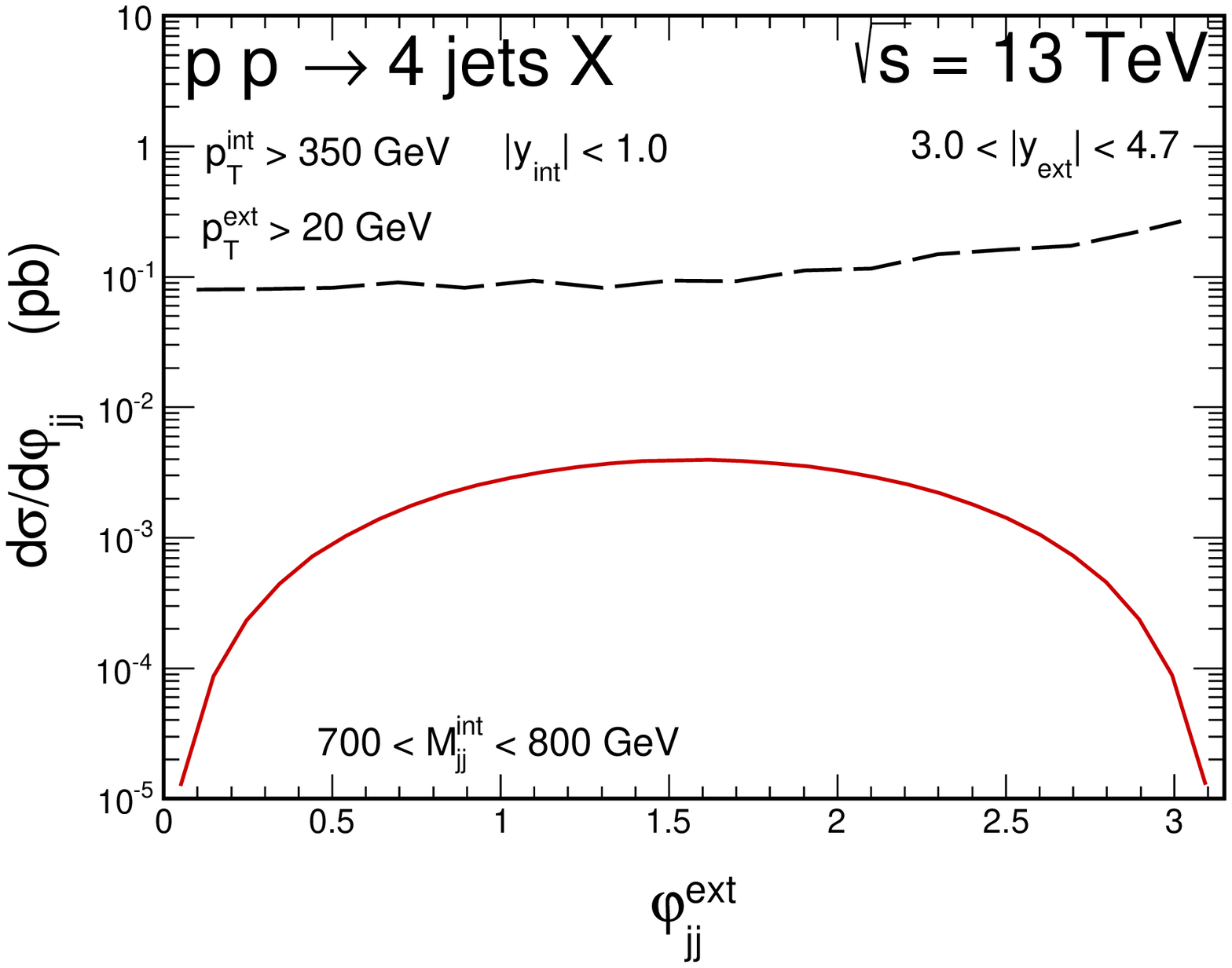}}
\end{minipage}
\hspace{0.5cm}
\begin{minipage}{0.47\textwidth}
 \centerline{\includegraphics[width=1.0\textwidth]{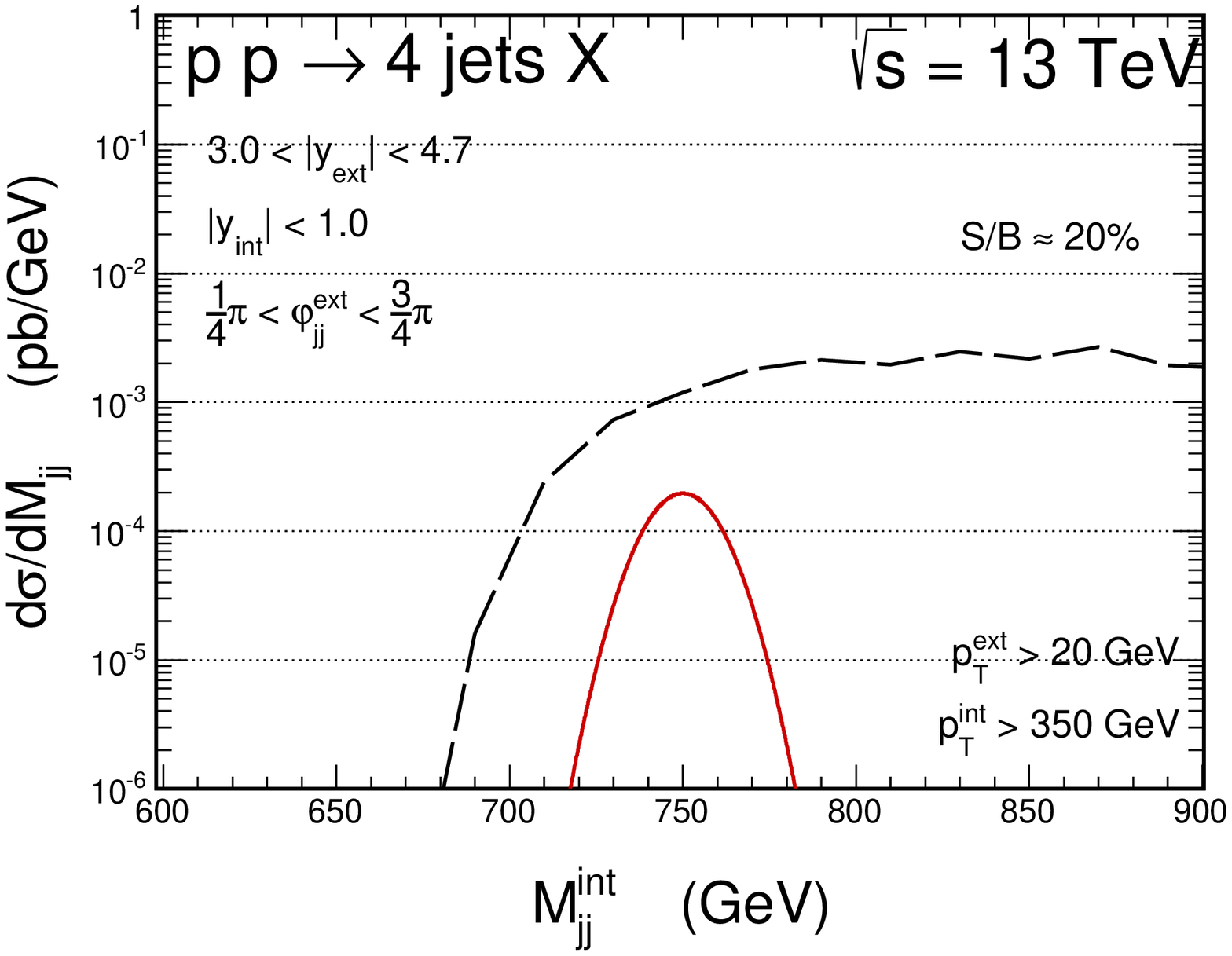}}
\end{minipage}
   \caption{
{\small Distribution in azimuthal angle between soft external jets
for the selected dijet invariant mass window, increased lower cut
on the transverse momenta of the internal jets and optimized rapidity cuts (left panel) and the dijet invariant mass
distribution with the specified cuts with extra limitation on the $\varphi^{ext}_{jj}$ variable (right panel).
 }
 }
 \label{fig:phi_cuts}
\end{figure}

Can we use some specific features of the fact that the resonance is
pseudoscalar? The answer is positive.
Similar problem was studied e.g. in Ref.~\cite{LLPS2016} for technipion and
in Ref.~\cite{Szczurek:2006bn} in the context
of exclusive production of $\eta$ meson in the $p p \to p p \eta$ reaction. 
One typical example is azimuthal correlations between the external 
spectator jets. The corresponding angular distribution in
azimuthal angle between the jets is shown in the left panel
of Fig.~\ref{fig:phi_cuts}. The signal has maximum at
$\phi_{jj}^{ext} \sim \pi/2$ (it would not be the case for scalar
resonance, see e.g. Ref.~\cite{Maciula:2010tv}).
The behaviour of the background is very different, it peaks at
$\phi_{jj}^{ext} \sim \pi$. Accepting only cases with 
$\phi_{jj}^{ext} \approx \pi/2$ would therefore improve the situation.
An example is shown in the right panel (see the detailed figure legend).
Now the signal-to-background ratio is about 0.2. 

Combining the above cuts may further improve the situation but 
the statistics may be lowered too much. One could return to 
the issue when better statistics will be accessible.

\begin{table}[tb]%
\caption{The calculated cross sections for the signal in femtobarns together with signal-to-background ratios and number of events for the present integrated luminosity of 3.2 fb$^{-1}$ at $\sqrt{s}=13$ TeV.}
\newcolumntype{Z}{>{\centering\arraybackslash}X}
\label{tab:cross sections}
\centering %
\begin{tabularx}{\linewidth}{c Z Z c }
\\[-4.ex] 
\toprule[0.1em] %
\\[-4.ex] 
\multirow{1}{6.5cm}{Kinematical cuts} & \multirow{1}{3.5cm}{Cross section [fb]} & \multirow{1}{1.5cm}{S/B [\%]} & \multirow{1}{2.cm}{Event rate} \\[+0.4ex]
\bottomrule[0.1em]
     
\multirow{1}{6.5cm}{$p_{T}^{ext} > 20$ GeV; $\; 2.5 < |y_{ext}| < 4.7$} &  \multirow{2}{3.5cm}{$69.00$}   & \multirow{2}{1.cm}{0.3}   & \multirow{2}{2.cm}{220} \\ [-0.2ex]
\multirow{1}{6.5cm}{$p_{T}^{int} > 100$ GeV; $\; |y_{int}| < 2.5$}            &                              &                                     &      \\ [+0.4ex]
\hline
\multirow{1}{6.5cm}{$p_{T}^{ext} > 20$ GeV; $\; 2.5 < |y_{ext}| < 4.7$} &  \multirow{2}{3.5cm}{$11.42$}   & \multirow{2}{1.cm}{4}   & \multirow{2}{2.cm}{36} \\ [-0.2ex]
\multirow{1}{6.5cm}{$p_{T}^{int} > 350$ GeV; $\; |y_{int}| < 2.5$}            &                              &                                     &      \\ [+0.4ex]
\hline
\multirow{1}{6.5cm}{$p_{T}^{ext} > 20$ GeV; $\; 3.0 < |y_{ext}| < 4.7$} &  \multirow{2}{3.5cm}{$6.13$}   & \multirow{2}{1.cm}{10}   & \multirow{2}{2.cm}{19} \\ [-0.2ex]
\multirow{1}{6.5cm}{$p_{T}^{int} > 350$ GeV; $\; |y_{int}| < 1.0$}            &                              &                                     &      \\ [+0.4ex]
\hline
\multirow{1}{6.5cm}{$p_{T}^{ext} > 20$ GeV; $\; 3.0 < |y_{ext}| < 4.7$} &  \multirow{3}{3.5cm}{$4.92$}   & \multirow{3}{1.cm}{20}   & \multirow{3}{2.cm}{15} \\ [-0.2ex]
\multirow{1}{6.5cm}{$p_{T}^{int} > 350$ GeV; $\;|y_{int}| < 1.0$}            &                              &                                     &      \\ [-0.2ex]
\multirow{1}{6.5cm}{$ \frac{1}{4}\pi < \varphi_{jj}^{ext} < \frac{3}{4}\pi $}            &                              &                                     &      \\ [+0.4ex]
\hline
\bottomrule[0.1em]

\end{tabularx}

\end{table}

The cross sections corresponding to different cuts are collected in
Table \ref{tab:cross sections}. 
If the resonance is really produced dominantly by the gluon-gluon
fusion, these cross sections are larger than those for the diphoton
channel. Therefore the four-jet study could be tried already
with the so-far recorded data.

\section{Conclusions}

The recently observed enhancement of the cross section at 750 GeV in the diphoton channel
may be produced by the gluon-gluon fusion. This would also mean
that it decays not only into the diphotons but also into 
two gluons (dijets).
As we have shown recently the observation of the resonance in the dijet
channel is practically not possible as the standard QCD background 
is fairly large compared to the signal.

As discussed recently, if the state is pseudoscalar then one can
expect large fraction of events with one or even two associated jets.
This could also mean that four-jet final state could be a better
choice in independent (to the diphoton channel) searches for the
signal of the so-far hypothetical 750 GeV state.

In the present studies we have focused on the analysis
of the four-jet final state in the context of searches for
the 750 GeV signal. Here, as an example, we have considered a walking technicolor model scenario.
In the case of the pseudoscalar state the corresponding amplitude 
is very specific.
Using its characteristic features we have analysed in detail
how to enhance the signal-to-background ratio. The $gg \to R(750)$ coupling constant was adjusted to reproduce the enhancement
in the diphoton channel. 
The four-jet background has been calculated with modern techniques implemented in the ALPGEN code.
Imposing specific cuts on jet transverse momenta and rapidities as well as
on azimuthal angle between external jets one can enhance the 
signal-to-background ratio up to about 20~\%.   
Unfortunately present statistics seems to be insufficient. Much better statistics will be available in a future.
The study of four dijet production and investigating dependence on cuts
may help in the future to assign a spin and parity to the new state, provided the signal is real.

\vspace{1cm}

{\bf Acknowledgments}

We are particularly indebted to Roman Pasechnik, Piotr Lebiedowicz
for a discussion on the issues discussed here.
The discussion with Shinya Matsuzaki on their version of walking
technipion model is acknowledged too.
This study was partially supported by 
the Polish National Science Center grant DEC-2014/15/B/ST2/02528 and 
by the Center for Innovation and
Transfer of Natural Sciences and Engineering Knowledge in
Rzesz{\'o}w.


\end{document}